\documentclass[twocolumn,showpacs,aps]{revtex4}

\usepackage[dvips,final]{graphics}
\usepackage{textcomp}
\usepackage{subeqnarray}

\begin{document}

\title{\center{Maximal Violation of Bell Inequalities using Continuous Variables Measurements}}

\author{J\'er\^ome Wenger, Mohammad Hafezi, Fr\'ed\'eric Grosshans, Rosa Tualle-Brouri and Philippe Grangier}
\affiliation{Laboratoire Charles Fabry de l'Institut  d'Optique, UMR 8501 du CNRS, F91403 Orsay - France\\
e-mail : philippe.grangier@iota.u-psud.fr}

\begin{abstract}
We propose a whole family of physical states that yield a
violation of the Bell CHSH inequality arbitrarily close to its
maximum value, when using quadrature phase homodyne detection.
This result is based on a new binning process called root binning,
that is used to transform the continuous variables measurements
into binary results needed for the tests of quantum mechanics
versus local realistic theories. A physical process in order to
produce such states is also suggested. The use of high-efficiency
spacelike separated homodyne detections with these states and this
binning process would result in a conclusive loophole-free test of
quantum mechanics.
\end{abstract}

\pacs{03.65.Ud, 03.67.-a, 42.50.Dv}

\maketitle

\section{Introduction}
Non-separability, or entanglement, has emerged as one of the most
striking feature of quantum mechanics. In 1935, it led Einstein,
Podolsky and Rosen to suggest \cite{EPR} that quantum mechanics is
incomplete, on the premise that any physical theory of nature must
be both \textquotedblleft local\textquotedblright and
\textquotedblleft realistic\textquotedblright. To quantify the
debate between quantum mechanics and local realistic (classical)
theories, Bell introduced a set of inequalities that must be
obeyed by any local realistic theory whereas they are vio\-lated
by quantum mechanics \cite{Bell,CHSH,CH}. These results shifted
the debate from the realm of philosophy to expe\-rimental physics.
The experiments done at the beginning of the 1980s by Aspect and
coworkers \cite{Aspect81,Aspect82a,Aspect82b} convin\-cingly
supported the predictions of quantum mechanics, but admittedly
left open two so-called  ``loopholes", that have to be addressed
for the evidence to be fully conclusive.

The first of these loopholes, called \textquotedblleft
locality\textquotedblright loophole, arises when the separation
between the measured states is not large enough to completely
discard the exchange of subluminal signals during the
measurements. The se\-cond loophole, called \textquotedblleft
detection-efficiency\textquotedblright loophole, occurs when the
particle detectors are inefficient enough so that the detected
events may be unrepresentative of the whole ensemble. In 1998,
Zeilinger \textit{et al.} \cite{Zeil} achieved communication-free
condition by using a type-II parametric down-conversion source and
fast random switching of the analyzers, that were separated by
about 400m. This closed the \textquotedblleft
locality\textquotedblright loophole, but their detection
efficiency was not sufficient to close the second loophole. In
2001, Rowe \textit{et al.} \cite{rowe} measured quantum
correlations bet\-ween two entangled beryllium ions with up to
80\% detection efficiency, closing the \textquotedblleft
detection-efficiency\textquotedblright loophole, but unfortunately
the ions were too close (about $3\mu m$) to avoid the
\textquotedblleft locality\textquotedblright loophole. Hence a
present challenge is to design and perform an experiment that
closes both loopholes to lead to a full logically consistent test
of any local realistic theory.

Quantum optics suggests good candidates, as photons can be
transported to sufficient long distances to avoid the
\textquotedblleft locality\textquotedblright loophole. To close
the \textquotedblleft detection-efficiency\textquotedblright
loophole, an alternative to photon-counting schemes consists in
quadrature-phase homodyne measurements, that use strong local
oscillators detected by highly efficient photodiodes. Up to date,
a few theoretical proposals that use quadrature-phase homodyne
detections have been made \cite{Reid,Gilchrist,Munro98,Munro99}
but for these set-ups the Bell inequality violation is a few
percents only, that lies far away of the maximal violation
attainable : $2\sqrt{2}$ (compared to a classical maximum of 2)
and $(1+\sqrt{2})/2$ (compared to a classical maximum of 1)
respectively for the Clauser-Horne-Shimony-Holt (CHSH)\cite{CHSH}
and Clauser-Horne (CH)\cite{CH} inequalities. Gilchrist \textit{et
al.} \cite{Reid,Gilchrist} use a \textit{circle} or pair coherent
state produced by non-degenerate parametric oscillation with the
pump adiabatically removed. This state leads to a theoretical
violation of about 1.015 ($>1$) of the CH Bell inequality. Munro
\cite{Munro99} considers correlated photon number states of the
form
\begin{equation} \label{PsiMunro}
|\Psi\rangle = \sum_{n=0}^N c_n |n\rangle|n\rangle
\end{equation}
where $N$ is truncated at $N=10$. He then performs a numerical
optimization on each $c_n$ coefficient to maximize the violation
of the CHSH Bell inequality when an homodyne measurement is
performed. For this specific state, the CHSH inequality is
violated by 2.076 ($>2$) and the CH inequality by 1.019 ($>1$).
From a different phase space approach, Auberson \textit{et al.}
\cite{Auberson} derive phase space Bell inequalities and propose a
state that yields a maximal violation of up to $2\sqrt{2}$ ($>2$).
This state can be expressed in the position space by
\begin{equation}
\Psi_{\pm}(q_1,q_2) = \frac{1}{2\sqrt{2}}\left[1\pm
   e^{i\frac{\pi}{4}} \textsf{sgn}(q_1) \textsf{sgn}(q_2)\right]f(|q_1|)f(|q_2|)
\end{equation}
where $f(q)$ is a regularized form of $1\over\sqrt{q}$, with
$\int_{-\infty}^{+\infty} dq\,f(q)^2=1$. The main problem with
this wave function lies in its singularities and phase switches.
Therefore, it requires nontrivial regularization procedures
to be considered as a suitable physical state.
Following these various attempts, we are thus looking for a ``simple"
physical state that would lead to a maximal violation of a Bell inequality.

\begin{figure}[b]
\center \includegraphics{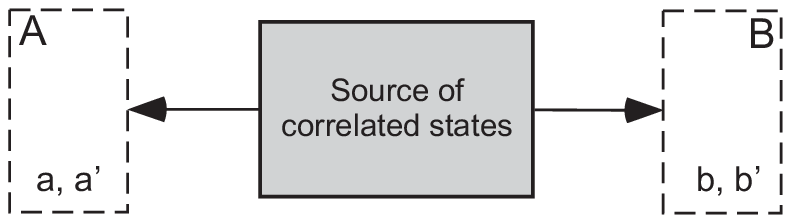} \caption{Schematic of a
generalized Bell experiment. The source generates correlated
states that are directed to the A and B devices used to perform
the measurements, with adjustable parameters a and b. Each
measurement provides a binary result \textquotedblleft
+\textquotedblright or \textquotedblleft -\textquotedblright
individually.} \label{expsetup}
\end{figure}

In this paper, we consider the Clauser-Horn-Shimony-Holt (CHSH)
Bell inequality \cite{CHSH}(sometimes referred to as the
\textit{spin} inequality). Fig.\ref{expsetup} depicts an idealized
setup for a general Bell inequality measurement. Two entangled
sub-states are viewed by two analyzers and detectors at locations
A and B, where a and b denote any adjustable parameter at A and B.
In our particular case, we will use quadrature-phase homodyne
measurements, which could have an efficiency high enough to close
the \textquotedblleft detection-efficiency\textquotedblright
loophole. Moreover, the apparatuses A and B can be in principle
spacelike separated, thereby excluding action at distance, and
closing also the \textquotedblleft locali\-ty\textquotedblright
loophole. We point out that in the present approach all the
detected light has to be taken into account, i.e. the relevant
signal is the photocurrent generated by the interferometric mixing
and photodetection of the local oscillator and input quantum
state. Therefore, no ``supplementary assumption"\cite{CHSH} will
be needed to interpret the data. Under these conditions, the CHSH
Bell inequality can be written \cite{CHSH}:
\begin{equation}
\label{ineq}
  S=|E(a',b')+E(a',b)+E(a,b')-E(a,b)|\leq2
\end{equation}
where the correlation function $E(a,b)$ is given by :
\begin{equation}
  E(a,b)=P_{++}(a,b)+P_{--}(a,b)-P_{+-}(a,b)-P_{-+}(a,b)
\end{equation}
with $P_{++}(a,b)$ the probability that a \textquotedblleft
+\textquotedblright occurs at both A and B, given $a$ and $b$.

In this article, we propose explicitly a set of physical states
that yield a violation of the CHSH inequality arbitrarily close to
its maximum value, when measured by an ideal quadrature-phase
homodyne detection. In section II, we describe how we convert the
continuous quadrature amplitude into a binary result
\textquotedblleft +\textquotedblright or \textquotedblleft
-\textquotedblright for each apparatus A,B using a process called
\textit{Root Binning}. In section III, a specific state that
yields a large violation of the CHSH inequality is presented. This
state is gene\-ralized in section IV to derive a whole family of
states that violate this Bell inequality. The issue of preparing
such states is addressed in section V, and various other
theoretical and practical issues are briefly discussed in the
conclusion.

\section{Root Binning}
To begin our study, we consider a state of the form of a
superposition of two two-particles wave functions with a relative
phase.
\begin{equation} \label{oestate}
|\Psi\rangle= \frac{1}{\sqrt{2}} \left(|f f \rangle+e^{i\theta}|g
g\rangle\right) \qquad 0 \leq \theta \leq 2\pi
\end{equation}
with $f$ real, even and normalized to unity while $g$ is assumed
real, odd and normalized to unity. This kind of state looks
similar to the one used by Auberson \textit{et al.}
\cite{Auberson}, but the $f$ and $g$ functions will be quite
different as well as the binning of the continuous variables being
measured.

The quadrature-phase homodyne measurement outputs a continuous
variable, and yet the majority of tests of local realism versus
quantum mechanics require a binary result. Hence for a given
quadrature measurement $q_i$ ($i=1,2$) at either location A or B,
we need to classify the result as either \textquotedblleft
+\textquotedblright or \textquotedblleft -\textquotedblright. In
ref. \cite{Auberson,Munro99,Gilchrist}, the \textquotedblleft
positive-negative\textquotedblright binning is used, that is the
result is classified as \textquotedblleft +\textquotedblright if
$q_i\geq0$ and \textquotedblleft -\textquotedblright if $q_i<0$.

The choice of binning is quite arbitrary. For the state
(\ref{oestate}) we can consider another type of binning, we call
\textit{root binning}, that depends on the roots of the functions
$f$ and $g$ (that are known in advance to the experimenters). We
assign \textquotedblleft +\textquotedblright when the result $q_i$
lies in an interval where $f(q)$ and $g(q)$ have the same sign,
and \textquotedblleft -\textquotedblright if $q_i$ is in an
interval where $f$ and $g$ have opposite signs. We define $D^+$ as
the union of the intervals in which $f(q)$ and $g(q)$ have the
same sign and $D^-$ as the union of the intervals in which $f(q)$
and $g(q)$ have opposite signs.
\begin{subeqnarray}
  D^+= \{  \forall ~ q ~ \epsilon ~\mathbf{ R } ~ | ~ f(q) ~
g(q)~ \geq ~ 0 \} \\
  D^-= \{ \forall ~ q ~ \epsilon ~ \mathbf{R} ~ | ~f(q)~ g(q) ~<
~0 \}
\end{subeqnarray}

Let us first consider the case when quadrature measurements in
position space have been performed on both sides. So the binary
probabilities we need for the CHSH type of Bell inequality will be
\begin{subeqnarray}
 P_{++}=\int_{D^+} \int_{D^+} dq_1 dq_2 P(q_1,q_2) \\
 P_{+-}=\int_{D^+} \int_{D^-} dq_1 dq_2 P(q_1,q_2) \\
 P_{-+}=\int_{D^-} \int_{D^+} dq_1dq_2 P(q_1,q_2) \\
 P_{--}=\int_{D^-} \int_{D^-} dq_1 dq_2 P(q_1,q_2)
\end{subeqnarray}
with
\begin{eqnarray}
  \lefteqn{P(q_1,q_2) = |\langle q_1| \langle q_2|\Psi\rangle|^2= } \nonumber \\
  & & \frac{1}{2}[f(q_1)^2f(q_2)^2+g(q_1)^2g(q_2)^2 \nonumber \\
  & & +2\ \cos\theta \ f(q_1)g(q_1)f(q_2)g(q_2)]
\end{eqnarray}

Hereafter, we calculate the correlation function for
(\ref{oestate}):
\begin{equation}
  E_{q_1,q_2}=P_{++}+P_{--}-P_{+-}-P_{-+}
\end{equation}
and as we have chosen $f$ even and $g$ odd, we get the remarkably
simple expression :
\begin{equation}
 E_{q_1,q_2}= V ^2\cos \theta
 \label{Exx}
\end{equation}
  where
\begin{eqnarray}
  V & = & \int_{D^+} f(q) g(q) dq - \int_{D^-} f(q) g(q) dq \nonumber \\
  & = & \int_{-\infty}^{+\infty} |f(q) g(q)| dq
\end{eqnarray}

A similar binning will be applied for the momentum part. Since we
suppose that $f(q)$ is a real and even function while $g(q)$ is
real and odd, $f(q)$ has a real even Fourier transform
$\tilde{f}(p)$ while $g(q)$ has an imaginary Fourier transform
$i \tilde{h}(p)$, where $\tilde{h}(p)$ is a real and odd
function. Using these properties,
and taking care of the supplementary $i$  factor, the same
reasoning applies for $\tilde{f}$ and $\tilde{h}$ as for $f$, $g$.
Denoting as $D'^+$ and $D'^-$ the intervals
associated with $\tilde{f}$ and $\tilde{h}$, we obtain
\begin{equation}
 E_{p_1,p_2}=- W ^2\cos \theta  \label{Epp}
 \end{equation}
where
\begin{eqnarray}
  W & = & \int_{D'^+} \tilde{f}(p) \tilde{h}(p) dp - \int_{D'^-}\tilde{f}(p) \tilde{h}(p) dp \nonumber\\
  & = & \int_{-\infty}^{+\infty}|\tilde{f}(p) \tilde{h}(p)| dp
\end{eqnarray}
and equivalently,
\begin{equation} E_{q_1,p_2}=- V ~W~\sin \theta   \label{Exp} \end{equation}
\begin{equation} E_{p_1,q_2}=- V ~W~\sin \theta   \label{Epx} \end{equation}
Hence by combining Eqs.(\ref{Exx}), (\ref{Epp}), (\ref{Exp}) and
(\ref{Epx}) we can write the CHSH inequality (\ref{ineq})
\begin{equation}
 S=| \cos(\theta) (V^2+W^2) - 2 \sin(\theta) V W|\leq 2.
 \end{equation}
The maximum of $S$ with respect to $\theta$ is obtained for
$\tan\theta_{m}=- 2VW /(V^2+W^2)$,
and we have $\theta_m\rightarrow -\pi/4$ as $V,W\rightarrow1$.
For this optimized $\theta_m$ we get the Bell inequality
\begin{equation}
\label{ineqeasy}
 S=| \sqrt{ W^4+V^4+6 V^2 W^2 } |\leq 2
\end{equation}
Using this really simple expression (\ref{ineqeasy}), the debate
bet\-ween quantum mechanics and local realistic theories boils
down to find functions $f$ and $g$ whose integrals $V$, $W$
violate (\ref{ineqeasy}). An interesting feature appears when the
distributions are eigenstates of the Fourier transform, so that
$V=W$ and equation (\ref{ineqeasy}) becomes
\begin{equation}
\label{ineqeasyTF}
 S=2\sqrt{2}~V^2\leq 2
\end{equation}
So if such functions have the right overlap needed to obtain
$V=1$, one will get the maximal violation of the above inequality,
which is obtained for $S=2\sqrt{2}$.

When compared to the positive-negative binning, root binning has
the advantage of having two parameters $V$ and $W$ to play with
while the positive-negative binning has only one \cite{Auberson}.
Moreover, as we will show now, the above Bell inequality is
violated by simple wave functions, that no longer have the
singularities that appeared in ref.\cite{Auberson}.

\section{Four paws Schr\"odinger cats}
In order to propose an explicit expression of a state that
violates the Bell inequality (\ref{ineqeasy}), let us first
consider the case of a Schr\"odinger cat state. The Schr\"odinger
cats, that are a superposition of two coherent states of
amplitudes $a$ and $-a$, involve intrinsic quantum features such
as negative Wigner functions, which make them interesting
candidates for our case (\ref{oestate}). We must choose an even
cat for $f(q)$ and an odd cat for $g(q)$ :
\begin{subeqnarray}
 f(q) \propto e^{-(q+a)^2/2}+e^{-(q-a)^2/2}\\
 g(q) \propto -e^{-(q+a)^2/2}+e^{-(q-a)^2/2}
 \end{subeqnarray}
unfortunately for this simple state we get $V=1$ and $W\simeq0.64$
for $a \rightarrow \infty$, so that $S\simeq1.90<2$. Therefore
this state cannot be used for violating Bell inequality (note that
Gilchrist \textit{et al.} \cite{Reid,Gilchrist} also consider
Schr\"odinger cats, but without getting a violation of Bell
inequalities).

\begin{figure}[t]
\center \includegraphics{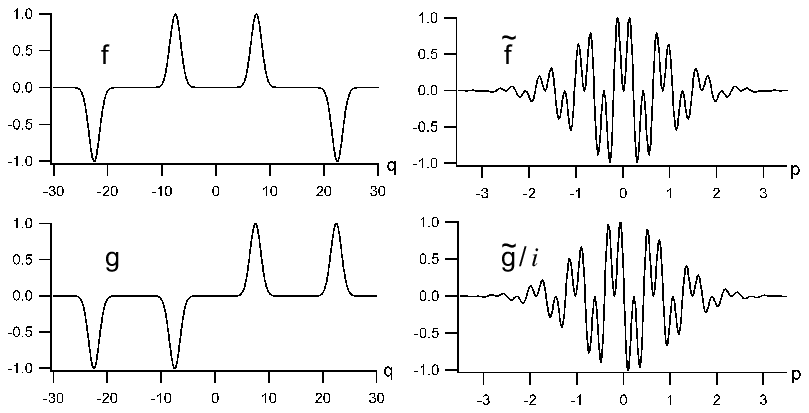} \caption{$f$ and $g$ for a
four-paws cat described by eqs.(\ref{f4}) in position space (left)
and momentum space (right), with $a=7.5$ ($\alpha=15$). Left axes
are in arbitrary units and normalized to unity.} \label{4Pfg}
\end{figure}

Instead of Schr\"odinger cats which can be alive or dead, we
consider acrobat cats which have 4 paws. Let us for instance
consider:
\begin{subeqnarray} \label{f4}
f(q)\propto-e^{-(q+3a)^2/2}+e^{-(q+a)^2/2}+e^{-(q-a)^2/2} \nonumber\\
~~~~~~~~~~~~-e^{-(q-3a)^2/2} \\
g(q)\propto-e^{-(q+3a)^2/2}-e^{-(q+a)^2/2}+e^{-(q-a)^2/2} \nonumber\\
~~~~~~~~~~~~+e^{-(q-3a)^2/2} \label{g4}
\end{subeqnarray}
$f(q)$ and $g(q)$ are depicted in Fig.\ref{4Pfg} together with
their Fourier Transform. Note that for this choice of functions,
each peak is distant from its neighbours of $\alpha=2a$, this
disposition yields an optimal overlap of $\tilde{f}$ and
$\tilde{g}$ and thus a high value of $S$. The best violation
appears when the peaks move off as $a \rightarrow \infty$. In that
case, $V=1$, $W=\frac{8}{3\pi}$ and thus we get the significant
violation of $S-2\simeq0.417$ (in facts, the condition $a
\rightarrow \infty$ appeared to be not so strict numerically, as
an amplitude $a=5$ is enough to obtain $S\simeq2.417$). Such a
violation represents a large improvement compared to Munro's best
result of 0.076 \cite{Munro99}, for a state with no singularity
and at least as easy to produce as Munro's $c_{n}$ optimized state
(\ref{PsiMunro}). However, we are still away from the maximal
value $2\sqrt{2}$ of the CHSH Bell inequality \cite{Cirelson}. In
the following section, we will propose a set of states to get
closer to the maximal violation.

\section{N paws Schr\"odinger cats} \label{Nsection}
The result obtained with the 4 paws cat suggests that a way to get
a stronger violation is to increase the total number $N$ of paws
of the cat states $|f_{N}\rangle$ and $|g_{N}\rangle$, with the
proper sign between the peaks. We thus define, for a given
amplitude $\alpha$
\begin{subeqnarray}\label{fN}
 f_{N;\alpha}(q)\propto \sum_{j=-\frac{N}{2}}^{\frac{N}{2}-1}\cos\left(\frac{\pi}{4}[2j+1]\right)
e^{-\frac{1}{2}\left(q-[j+\frac{1}{2}]\alpha\right)^2}   \\
\slabel{gN} g_{N;\alpha}(q)\propto
\sum_{j=-\frac{N}{2}}^{\frac{N}{2}-1}\sin\left(\frac{\pi}{4}[2j+1]\right)
e^{-\frac{1}{2}\left(q-[j+\frac{1}{2}]\alpha\right)^2}
\end{subeqnarray}

\begin{table}[b]
\begin{tabular}{|c|c|c|c|c|c|c|}
  \hline
  \textit{N} & 2 & 4 & 6 & 8 & 10 & 12 \\ \hline
  \textit{S} & 1.895 & 2.417 & 2.529 & 2.611 & 2.649 & 2.681 \\ \hline
\end{tabular}
  \caption{\textit{S} for N-paws cat defined by eqs.(\ref{fN}) and $\alpha=15$, each peak having the same high.} \label{tabNS}
\end{table}

\begin{figure}[t]
\center \includegraphics{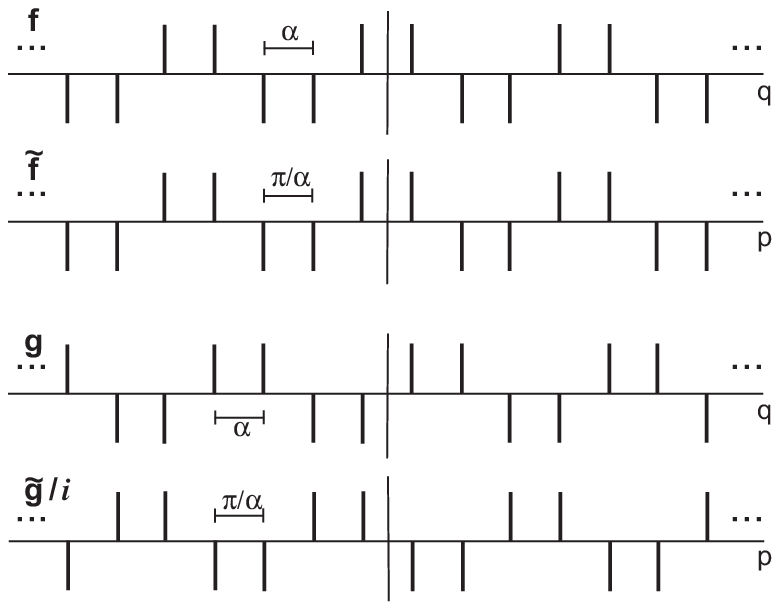} \caption {Infinite-paws states,
represented in the position and momentum phase space. Thick
segments denote Dirac Delta functions.} \label{infPfg}
\end{figure}

Table \ref{tabNS} presents the results of the calculation of $S$
according to the formula (\ref{ineqeasy}) for the state defined by
(\ref{oestate}) and (\ref{fN}). As expected, the quantity $S$
increases with the number of paws and tends to $2\sqrt{2}$. To
prove this point, let us consider the two following distributions
that have an infinite number of paws. These distributions are
depicted with their Fourier Transform in fig.\ref{infPfg}.
\begin{subeqnarray}\label{finf}
  f_{\infty;\alpha}(q)\propto \sum_{j=-\infty}^{+\infty}\delta(q-\alpha (j+1/2))\
\cos\left(\frac{\pi q}{2\alpha}\right) \\
  g_{\infty;\alpha}(q)\propto \sum_{j=-\infty}^{+\infty}\delta(q-\alpha (j+1/2))\
\sin\left(\frac{\pi q}{2\alpha}\right) \label{ginf}
\end{subeqnarray}

Up to scaling factors, these distributions appear to be almost
identical to their Fourier transform, while for the specific
amplitude $\alpha=\sqrt{\pi}$ they are exact eigenstates of the
Fourier transform. Moreover, as $f_{\infty}$ and $g_{\infty}$
overlap perfectly, these distributions would yield the maximal
vio\-lation of the CHSH inequality, as $S=2\sqrt{2}$ from
eq.(\ref{ineqeasyTF}). Of course, they are unphysical
(non-normalizable) states, but the wave functions $f_{N}$ and
$g_{N}$ in (\ref{fN}) can be considered as regularized forms of
$f_{\infty}$ and $g_{\infty}$, widening the Dirac Delta functions
to Gaussians and taking a finite number of paws. Thus one can
understand that $S\rightarrow2\sqrt{2}$ as $N\rightarrow\infty$.

Another regularization of the wave functions (\ref{finf}) consists
in a widening of the Dirac functions to Gaussians of width $s$
associated with a Gaussian envelope of width $1/s$:
\begin{subeqnarray} \label{fgauss}
  f_{\infty; \alpha,s}(q)\propto G_{1/s}(q)[f_{\infty; \alpha}*G_{s}(q)] \\
  g_{\infty; \alpha,s}(q)\propto G_{1/s}(q)[g_{\infty;
  \alpha}*G_{s}(q)]\label{ggauss}
\end{subeqnarray}
where $G_{s}(q)=\exp(-\frac{q^{2}}{2s^{2}})$, $*$ denotes the
convolution product, and $s$ is a squeezing parameter. When $s\ll
1$, one indeed has:
\begin{subeqnarray}
  \widetilde{f}_{\infty; \alpha,s}(p)\propto G_{s}*[f_{\infty; \pi/\alpha}G_{1/s}](p)\approx
  f_{\infty;\pi/\alpha,s}(p) \\
  \widetilde{g}_{\infty; \alpha,s}(p)\propto G_{s}*[g_{\infty; \pi/\alpha}G_{1/s}](p)\approx g_{\infty;\pi/\alpha,s}(p)
\end{subeqnarray}
The violation exhibits a symmetry, as
$S(\alpha)=S(\frac{\pi}{\alpha})$, with a maximum reached for
$\alpha=\sqrt{\pi}$ when $f$ and $g$ are approximately eigenstates
of the Fourier transform.

\begin{figure}[h]
\center \includegraphics{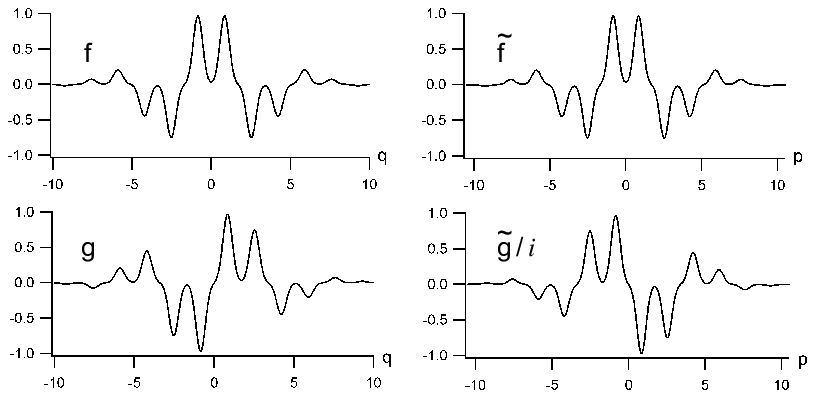} \caption {$N=12$ cat
states described by eqs(\ref{fgauss}) presented in position space
(left) and in momentum space (right) for parameters $\alpha=1.8$
and $s=0.3$. Left axes are in arbitrary units and normalized to
unity.} \label{12Pfg}
\end{figure}

Thanks to the Gaussian envelope, the above functions can be
truncated to a finite total number of paws $N$ without affecting
numerically the $f$ and $g$ functions, provided:
\begin{equation} \label{Nlimitinf}
  N>\frac{2\sqrt{2|\ln\varepsilon|}}{\alpha s}
\end{equation}
where $\varepsilon$ is an arbitrary small tolerance parameter. For
$\varepsilon=0.01$, $\alpha=\sqrt{\pi}$ and $s=0.3$, the condition
yields $N\geq12$. Given these parameters and $N=12$, we get
$S\approx2\sqrt{2}$ with a relative error of 0.01\%. Regardless to
this condition, one may also arbitrarily choose to limit the above
functions to $N$ paws for given parameters $s$ and $\alpha$, but
as such $f$ and $g$ can not directly be considered as truncated
$f_{\infty; \alpha,s}$ and $g_{\infty; \alpha,s}$, the best
amplitude will differ from $\sqrt{\pi}$ and $S$ will slightly move
away from $2\sqrt{2}$. Some results obtained for $s=0.3$ are
presented on table \ref{tabNSopt} for $N$ running from 4 to 12.
The value of $\alpha$ used in this table is numerically calculated
in order to maximize $S$. Here the violation is considerably
improved compared to table \ref{tabNS}, with for instance
$S=2.764$ with $N=4$, and $S=2.828$ with $N=10$. In figure
\ref{12Pfg} we display $f(q)$ and $g(q)$ for the case $N=12$
showing that these functions are nearby self Fourier Transform.

In appendix \ref{appendix1}, we rewrite these states on the
Fock-states basis and express the $f$ and $g$ functions as
combinations of Hermite polynomials.

\begin{table}[h]
\begin{tabular}{|c|c|c|c|c|c|}
  \hline
  \textit{N} & 4 & 6 & 8 & 10 & 12 \\ \hline
  $\alpha_{opt}$ & 2.6 & 2.3 & 2 & 1.8 & 1.8 \\ \hline
  \textit{S} & 2.764 & 2.823 & 2.826 & 2.828 & 2.828 \\ \hline
\end{tabular}
  \caption{\textit{S} as a function of the number of peaks \textit{N} for a squeezing parameter \textit{s}=0.3
and a gaussian envelope of width \textit{1/s}, for an optimized
amplitude $\alpha_{opt}$} \label{tabNSopt}
\end{table}

\section{Conditional preparation of entangled N paws Schr\"odinger cats}
Preparing entangled $N$ paws Schr\"odinger cats is a challenging
task, we begin by focusing our attention on how one single $N$
paws Schr\"odinger cat as defined by (\ref{gN}) could be
generated. This state has strong similari\-ties with the
\textit{encoded states} introduced by Gottesman and coworkers
\cite{Gottesman} to perform quantum error-correction codes.
Recently, Travaglione and Milburn \cite{Milburn} presented a
proposal to generate non-deterministically such \textit{encoded
states}. Following this study, we will first show how to generate
the state $|g\rangle$ by applying a specific sequence of
operations similar to \cite{Milburn} and then derive a setup to
produce the whole state (\ref{oestate}).

The preparation procedure begins with the quantum system in the
vacuum state $|0\rangle$ and an ancilla qubit in the ground state
$|0\rangle_a$ (if some squeezing parameter is needed, one may take
 a squeezed vacuum as quantum system and follow the procedure we describe here. For
simplicity reasons, we set $s=1$). Let us first apply:
\begin{equation}
  H~e^{- i \alpha p \sigma_z}~H
\end{equation}
with $p$ is the momentum operator applied to the continuous
variable state, $H$ the Hadamard gate and $\sigma_z$ the Pauli
matrix applied to the qubit :
\begin{equation}
  H=\frac{1}{\sqrt{2}}\left(\begin{array}{cc} 1 & 1 \\
1 & -1 \end{array}\right) ~~~~~~~~\sigma_z=\left(\begin{array}{cc} 1 & 0 \\
0 & -1 \end{array}\right)
\end{equation}
We then have a probability 1/2 of measuring the qubit either in
the excited or in the ground state. If it is found in the
$|1\rangle_a$ state, the continuous variable is left in the state
$|\Upsilon_1\rangle\propto -|-\alpha\rangle + |\alpha\rangle$,
otherwise the procedure is stopped and we try again. The qubit is
then bit-flipped to $|0\rangle_a$ and we go on by applying the
sequence
\begin{equation}
  H~e^{- i 2\alpha p \sigma_z}~H
\end{equation}
Measuring the qubit in the $|0\rangle_a$ results in the continuous
variable left in the state $|\Upsilon_2\rangle\propto
-|-3\alpha\rangle+|-\alpha\rangle-|\alpha\rangle+|3\alpha\rangle$.
To increase the number of paws, we iterate the following procedure
given $|\Upsilon_{n-1}\rangle$ and the qubit in $|0\rangle_a$ :
\begin{itemize}
\item Apply the operators
\begin{equation} \label{iter}
  H~e^{- i 2^{n-1}\alpha p \sigma_z}~H
\end{equation}
\item Measure the qubit
\item If the qubit was in the state $|0\rangle_a$, we have
  created $|\Upsilon_{n}\rangle$
\item If the qubit was in the state $|1\rangle_a$, discard
  and try again.
\end{itemize}

Once the number of paws is considered satisfactory, we stop the
previous iteration. The last point to generate $|g\rangle$ is to
apply the following sequence
\begin{equation}
  H~e^{- i \frac{\alpha}{2} p \sigma_z}~H
\end{equation}
If the qubit is found in the state $|0\rangle_a$, we have created
the state $|g\rangle$ as defined in eq.(\ref{gN}). If the protocol
was stopped after $n$ iterations (corresponding to
$|\Upsilon_{n}\rangle$), the state $|g\rangle$ is created with
probability $1/2^{n+1}$ and shows $N=2^{n+1}$ paws. In the last
section of ref.\cite{Milburn}, Travaglione and Milburn briefly
consider the question of the physical implementation of this
iteration process using a radio-frequency ion trap. We refer the
reader to this article for further details.

\begin{figure}[t]
\center \includegraphics{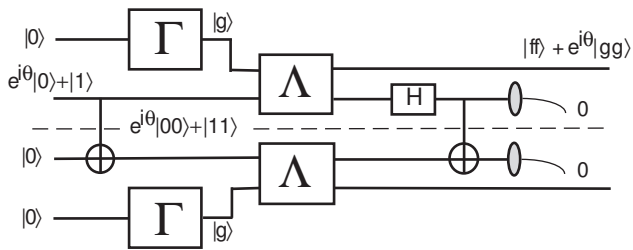} \caption {Schematic of the
setup used to generate the state $|\Psi\rangle= \frac{1}{\sqrt{2}}
\left(|f f \rangle+e^{i\theta}|g g\rangle\right)$ with $f$ and $g$
defined by eqs.(\ref{fN}). See text for the notations.}
\label{prepa}
\end{figure}

For the creation process of the state $|\Psi\rangle=
\frac{1}{\sqrt{2}} \left(|f f \rangle+e^{i\theta}|g
g\rangle\right)$, we present on fig.\ref{prepa} a global view of
our scheme. The $\Gamma$-labelled box corresponds to the
generation of $|g\rangle$ we just saw. Our process is based on a
CNOT gate that entangles two qubits
($|\psi\rangle_{a}=\frac{1}{\sqrt{2}}(|11\rangle_{a}+e^{i\theta}|00\rangle_{a})$).
Each qubit is then associated with a state $|g\rangle$ through the
following operator
\begin{equation}
  \Lambda=e^{\frac{i\pi}{4}\left(\frac{q}{\alpha/2}-1\right)\left(1-\sigma_z\right)}
\end{equation}
where $q$ is the position operator of the $|g\rangle$ state and
$\sigma_z$ the Pauli matrix applied to the qubit. If the qubit is
at zero, the $|g\rangle$ state will be left unchanged. If the
qubit is at one, for $s\ll\alpha$ the sign of each two peak is
changed so that the $|g\rangle$ state is changed into the
$|f\rangle$ state. After this operator $\Lambda$ has been applied,
the whole state is left in
\begin{equation}
    e^{i\theta}~|g~g~0~0\rangle + |f~f~1~1\rangle
\end{equation}
In order to disentangle the two qubits from the conti\-nuous part,
one has to make the qubits pass through an Hadamard plus a CNOT
gate. Finally measuring one qubit in the zero state will project
the continuous variable system onto the awaited state
\begin{equation}
|\Psi\rangle= \frac{1}{\sqrt{2}} \left(|f f \rangle+e^{i\theta}|g
g\rangle\right)
\end{equation}

Such a process generates the N-paws Schr\"odinger cats as defined
in (\ref{fN}) where each peak has the same height. The case
(\ref{fgauss}) with a gaussian envelope is more complicated to
produce because each peak must be separately weighted, but as seen
in table \ref{tabNS}, the violation for a state (\ref{fN}) with
equal heights is already quite strong ($S\simeq2.41$ for $N=4$).

\section{Conclusion}
Considering quadrature phase homodyne detection, we have derived a
new binning process called \textit{root binning} to transform the
continuous variables measured into binary results to be used in
the test of quantum mechanics versus local realistic theories. For
this process, we propose a whole family of physical states that
yield a violation arbitrarily close to the maximal violation in
quantum mechanics and much stronger than previous works in the
domain \cite{Reid, Gilchrist, Munro98, Munro99}.

We have also tested root binning on other interesting forms of
Bell inequality that are the information-theoretic inequalities
developed by Braunstein and Caves \cite{Braustein} and generalized
by Cerf and Adami \cite{Cerf}. Using quadrature measurements, root
binning and our state defined by eqs.(\ref{oestate},\ref{fgauss}),
we unfortunately could find no violation for neither Braunstein's
nor Cerf's form of information-theoretic Bell inequality. Our
state in fact tends to the minimum limit for violation of these
information inequalities when $V\rightarrow1$ and $W\rightarrow1$.
As a matter of fact, the binning process discards a lot of
information that lies within each interval of the binning. This
information loss may prevent any violation of
information-theoretic inequalities.

As a conclusion, let us point out that though the present idea
sounds quite attractive, its practical implementation is very
far-fetched. Though we do propose a theoretical scheme to prepare
the required states, it relies on various features (coupling
Hamiltonian, CNOT gates) that are not presently available with the
required degree of efficiency. For going in more details into the
implementation of the present scheme, inefficiencies and
associated decoherence effects should be examined in detail for
each required step, i.e. preparation, propagation, and detection.
Also, various possible implementations of the proposed scheme
should be considered \cite{rowe,Milburn,Brune,Lvovsky}.

Such a study is out of the scope of the present paper, that has
mostly the goal to show that arbitrarily high violations of Bell
inequalities are in principle possible, by using
continuous-variables measurements and physically meaningful -
though hardly feasible - quantum states.

This work is supported by the European IST/FET/QIPC
program, and by the French programs \textquotedblleft ACI
Photonique\textquotedblright and \textquotedblleft
ASTRE\textquotedblright.


\appendix

\section{Expression in the Fock basis}
An interesting point is to study the decomposition of states
$|f\rangle$ and $|g\rangle$ on the Fock basis $|n\rangle$, which
are experimentally accessible. Starting from states (\ref{fgauss})
with $\alpha=\sqrt{\pi}$, $s=0.4$ and with $N$ satisfying
condition (\ref{Nlimitinf}), we obtained after truncating at the
$14^{th}$ order:
\begin{subeqnarray}
|f\rangle =
\sqrt{0.459}|0\rangle-\sqrt{0.491}|4\rangle-\sqrt{0.008}|8\rangle \nonumber\\
~~~~~~~~~~~~~~~~~~~~~~~~~~~~~~~~-\sqrt{0.042}|12\rangle\\
|g\rangle
=\sqrt{0.729}|1\rangle+\sqrt{0.155}|5\rangle-\sqrt{0.107}|9\rangle \nonumber\\
~~~~~~~~~~~~~~~~~~~~~~~~~~~~~~~~-\sqrt{0.009}|13\rangle
\end{subeqnarray}
These states allow to reach a violation $S-2=0.81$. The state
$|f\rangle$ only involves orders $n\equiv0~(mod~4)$, as we have
$n\equiv1~(mod~4)$ for $|g\rangle$. This directly comes from the
fact that $\langle p|n\rangle=(-i)^{n}\langle q|n\rangle_{q=p}$
and $(-i)^{4}=1$, so that states of the form $\sum_{n\equiv a~
(mod~4)}c_{n}|n\rangle$ are eigenvectors of the Fourier transform.
From these considerations we have obtained $S=2.68$ for:
\begin{subeqnarray}
|f\rangle =\sqrt{0.585}|0\rangle-\sqrt{0.415}|4\rangle\\
|g\rangle =\sqrt{0.848}|1\rangle+\sqrt{0.152}|5\rangle
\end{subeqnarray}
and $S=2.3$ for:
\begin{equation}
|f\rangle =\sqrt{0.67}|0\rangle-\sqrt{0.33}|4\rangle, \; \; \; |g\rangle
=|1\rangle
\end{equation}
These states are quite simple, and it is expected that a specific
procedure to produce them might be designed. \label{appendix1}



\end{document}